\documentstyle[aps]{revtex}

\begin{document}

\title{Helicity Basis for Spin 1/2 and 1}

\author{Valeri V. Dvoeglazov and J. L. Quintanar Gonz\'alez}

\address{Universidad de Zacatecas\\ A. P. 636, Suc. UAZ, C. P. 98062, Zacatecas, Zac., M\'exico\\ 
E-mail: valeri@ahobon.reduaz.mx, el\_leo\_xyz@yahoo.com.mx}

\maketitle

\begin{abstract}
We study the theory of the $(1/2,0)\oplus (0,1/2)$ and $(1,0)\oplus (0,1)$
representations in  the helicity basis. The helicity eigenstates
are {\it not} the parity eigenstates. This is in accordance with the
idea of Berestetski\u{\i}, Lifshitz and Pitaevski\u{\i}. The behaviour
of the helicity eigenstates with respect to the charge conjugation, $CP$- conjugation is also different comparing with the parity eigenstates.
\end{abstract}
{}{}

\bigskip
\bigskip

\section{Introduction.}

Recently we generalized the Dirac formalism~\cite{Barut,DasG,Dv2,Dv2a} and
the Bargmann-Wigner formalism~\cite{Dv3,Dv4,Dv5}. On this basis we proposed a set of  equations for antisymmetric tensor (AST) field. Some of them imply
parity-violating transitions. In this paper we are going to study
transformations from the standard basis
to the helicity basis in the Dirac theory and in the $(1,0)\oplus (0,1)$ Sankaranarayanan-Good theory~\cite{SG,Dva}. The spin basis rotation
{\it changes} properties of the corresponding states with respect to
parity. The parity is a physical quantum number; so, we try to extract
corresponding physical contents from considerations of the various spin
bases.

\section{The $(1/2,0)\oplus (0,1/2)$ case.}

Beginning the consideration of the helicity basis, we observe that
it is well known that the operator $\hat {\bf S}_3 = {\bbox
\sigma}_3/2\otimes I_2$ does not commute with the Dirac Hamiltonian unless
the 3-momentum is aligned along with the third axis and the plane-wave
expansion is used:
\begin{equation}
[\hat{\cal H},\hat {\bf S}_3]_- = (\gamma^0{\bbox\gamma}^k \times
{\bf\nabla}_i )_3
\end{equation}
Moreover, Berestetski\u{\i}, Lifshitz and
Pitaevski\u{\i} wrote~\cite{Lan}: ``... the orbital angular momentum ${\bf
l}$ and the spin ${\bf s}$ of a moving particle are not separately
conserved. Only the total angular momentum ${\bf j}= {\bf l}+{\bf s}$ is
conserved. The component of the spin in any fixed direction (taken as
$z$-axis is therefore also not conserved, and cannot be used to enumerate
the polarization (spin) states of moving particle." The similar conclusion has
been given by Novozhilov in his book~\cite{Novozh}. On the other hand, the helicity operator ${\bbox\sigma}\cdot
\widehat {\bf p}/2 \otimes I$, $\widehat {\bf p} = {\bf p}/\vert {\bf
p}\vert$, commutes with the Hamiltonian (more precisely, the commutator is
equal to zero when acting on the one-particle plane-wave solutions).

So, it is a bit surprising, why the 4-spinors have been studied so well
when the basis have been chosen in such a way that they were eigenstates
of the $\hat {\bf S}_3$ operator:
\begin{eqnarray}
u_{{1\over 2},{1\over 2}} = N_{1\over 2}^+\pmatrix{1\cr0\cr1\cr0\cr}\,,\,
u_{{1\over 2},-{1\over 2}} =N_{-{1\over 2}}^+ \pmatrix{0\cr1\cr0\cr1\cr}\,,\,
v_{{1\over 2},{1\over 2}} = N_{1\over 2}^-\pmatrix{1\cr0\cr-1\cr0\cr}\,,\,
v_{{1\over 2},-{1\over 2}} =N_{-{1\over 2}}^-
\pmatrix{0\cr1\cr0\cr-1\cr}\,,\label{sb1}
\end{eqnarray}
and, oppositely, the
helicity basis case has not been studied almost at all
(see, however, refs.~\cite{Novozh,Grei,JW}.  Let me remind
that the boosted 4-spinors in the `common-used' basis are
\begin{eqnarray}
u_{{1\over 2},{1\over 2}} = {N_{1\over
2}^+\over \sqrt{2m (E+m)}}
\pmatrix{p^++m\cr p_r\cr p^- +m\cr -p_r\cr}\,,\,
u_{{1\over 2},-{1\over 2}} ={N_{-{1\over 2}}^+
\over \sqrt{2m (E+m)}}\pmatrix{p_l\cr p^-+m\cr -p_l\cr p^++m\cr}\,, \\
v_{{1\over 2},{1\over 2}} = {N_{1\over
2}^- \over \sqrt{2m (E+m)}}\pmatrix{p^++m\cr p_r\cr -p^- -m\cr p_r\cr}\,,\,
v_{{1\over 2},-{1\over 2}} ={N_{-{1\over
2}}^- \over \sqrt{2m (E+m)}}\pmatrix{p_l\cr p^-+m\cr p_l\cr -p^+-m\cr}\,,
\end{eqnarray} 
$p^\pm = E\pm p_z$, $p_{r,l}= p_x\pm ip_y$.
They  are the parity eigenstates with the eigenvalues of $\pm 1$. The matrix $\gamma_0=\pmatrix{0&\openone\cr \openone &
0\cr}$ is used in the parity operator.

Let me turn now your attention to the helicity spin basis.
The 2-eigenspinors of the helicity operator 
\begin{eqnarray}
{1\over 2} {\bbox \sigma}\cdot\widehat
{\bf p} = {1\over 2} \pmatrix{\cos\theta & \sin\theta e^{-i\phi}\cr
\sin\theta e^{+i\phi} & - \cos\theta\cr}
\end{eqnarray}
can be defined as follows~\cite{Var,Dv1}:
\begin{eqnarray}
\phi_{{1\over 2}\uparrow}=\pmatrix{\cos{\theta \over 2} e^{-i\phi/2}\cr
\sin{\theta \over 2} e^{+i\phi/2}\cr}\,,\quad
\phi_{{1\over 2}\downarrow}=\pmatrix{\sin{\theta \over 2} e^{-i\phi/2}\cr
-\cos{\theta \over 2} e^{+i\phi/2}\cr}\,,\quad\label{ds}
\end{eqnarray}
for $\pm 1/2$ eigenvalues, respectively.

We start from the Klein-Gordon equation, generalized for
describing the spin-1/2  particles (i.~e., two degrees
of freedom); $c=\hbar=1$:\footnote{The following method is due to the van der Waerden, Sakurai and Gersten
works, see ref.~\cite{Gerst}.}
\begin{equation}
(E+{\bbox \sigma}\cdot {\bf p}) (E- {\bbox \sigma}\cdot {\bf p}) \phi
= m^2 \phi\,.\label{de}
\end{equation}
It can be re-written in the form of the set of two first-order equations
for 2-spinors. Simultaneously, we observe that they may be chosen
as eigenstates of the helicity operator which present
in (\ref{de}):\footnote{This opposes to the choice of the basis
(\ref{sb1}), where 4-spinors are the eigenstates of the parity operator.}
\begin{eqnarray}
(E-({\bbox\sigma}\cdot {\bf p})) \phi_\uparrow &=& (E-p) \phi_\uparrow
=m\chi_\uparrow \,,\\
(E+({\bbox\sigma}\cdot {\bf p})) \chi_\uparrow &=& (E+p) \chi_\uparrow
=m\phi_\uparrow \,,
\end{eqnarray}
and
\begin{eqnarray}
(E-({\bbox\sigma}\cdot {\bf p})) \phi_\downarrow &=& (E+p) \phi_\downarrow
=m\chi_\downarrow\,, \\
(E+({\bbox\sigma}\cdot {\bf p})) \chi_\downarrow &=& (E-p) \chi_\downarrow
=m\phi_\downarrow \,.
\end{eqnarray}
If the $\phi$ spinors are defined by the equation (\ref{ds}), then we
can construct the corresponding $u-$ and $v-$ 4-spinors:
\begin{eqnarray}
u_\uparrow &=&
N_\uparrow^+ \pmatrix{\phi_\uparrow\cr {E-p\over m}\phi_\uparrow\cr} =
{1\over \sqrt{2}}\pmatrix{\sqrt{{E+p\over m}} \phi_\uparrow\cr
\sqrt{{m\over E+p}} \phi_\uparrow\cr}\,,
u_\downarrow = N_\downarrow^+ \pmatrix{\phi_\downarrow\cr
{E+p\over m}\phi_\downarrow\cr} = {1\over
\sqrt{2}}\pmatrix{\sqrt{{m\over E+p}} \phi_\downarrow\cr \sqrt{{E+p\over
m}} \phi_\downarrow\cr}\,,\label{s1}\\
v_\uparrow &=& N_\uparrow^- \pmatrix{\phi_\uparrow\cr
-{E-p\over m}\phi_\uparrow\cr} = {1\over \sqrt{2}}\pmatrix{\sqrt{{E+p\over
m}} \phi_\uparrow\cr
-\sqrt{{m\over E+p}} \phi_\uparrow\cr}\,,
v_\downarrow = N_\downarrow^- \pmatrix{\phi_\downarrow\cr
-{E+p\over m}\phi_\downarrow\cr} = {1\over
\sqrt{2}}\pmatrix{\sqrt{{m\over E+p}} \phi_\downarrow\cr -\sqrt{{E+p\over
m}} \phi_\downarrow\cr}\,,\label{s2}
\end{eqnarray} 
where the normalization to the unit
($\pm 1$) was used:\footnote{Of course, there are no any mathematical
difficulties
to change it  to the normalization to $\pm m$, which may be more convenient
for the study of the massless limit.}
\begin{eqnarray}
\bar u_\lambda u_{\lambda^\prime} &=& \delta_{\lambda\lambda^\prime}\,,
\bar v_\lambda v_{\lambda^\prime} = -\delta_{\lambda\lambda^\prime}\,,\\
\bar u_\lambda v_{\lambda^\prime} &=& 0 =
\bar v_\lambda u_{\lambda^\prime}
\end{eqnarray}
One can prove that the matrix
\begin{equation}
P=\gamma^0 = \pmatrix{0&\openone\cr\openone & 0\cr}
\label{par}
\end{equation}
can also be used in the parity operator as well as
in the original Dirac basis. Indeed, the 4-spinors
(\ref{s1},\ref{s2}) satisfy the Dirac equation in the spinorial
representation of the $\gamma$-matrices (see straightforwardly
from (\ref{de})). Hence, the parity-transformed function
$\Psi^\prime (t, -{\bf x})=P\Psi (t,{\bf x})$ must satisfy
\begin{equation}
[i\gamma^\mu \partial_\mu^{\,\prime} -m ] \Psi^\prime (t,-{\bf x}) =0 \,,
\end{equation}
with $\partial_\mu^{\,\prime} = (\partial/\partial t, -{\bf \nabla}_i)$.
This is possible when $P^{-1}\gamma^0 P = \gamma^0$ and
$P^{-1} \gamma^i P = -\gamma^i$. The matrix (\ref{par})
satisfies these requirements, as in the textbook case.

Next, it is easy to prove that one can form the projection
operators
\begin{eqnarray}
P_+ &=&+\sum_{\lambda} u_\lambda ({\bf p}) \bar u_\lambda ({\bf p})
=\frac{p_\mu \gamma^\mu +m}{2m}\,,\\
P_- &=&-\sum_{\lambda} v_\lambda ({\bf p}) \bar v_\lambda ({\bf p})
= \frac{m- p_\mu \gamma^\mu}{2m}\,,
\end{eqnarray}
with the properties $P_+ +P_- =1$ and $P_\pm^2 = P_\pm$.
This permits us to expand the 4-spinors defined in the basis  (\ref{sb1})
in  linear superpositions of the helicity basis
4-spinors and to find corresponding coefficients of the expansion:
\begin{eqnarray}
u_\sigma ({\bf p}) &=& A_{\sigma\lambda} u_\lambda ({\bf p})
+ B_{\sigma\lambda} v_\lambda ({\bf p})\,,\\
v_\sigma ({\bf p}) &=& C_{\sigma\lambda} u_\lambda ({\bf p})
+ D_{\sigma\lambda} v_\lambda ({\bf p})\,.
\end{eqnarray}
Multiplying the above equations by $\bar u_{\lambda^\prime}$,
$\bar v_{\lambda^\prime}$ and using the normalization conditions,
we obtain $A_{\sigma\lambda}= D_{\sigma\lambda}= \bar u_\lambda u_\sigma$,
$B_{\sigma\lambda}=C_{\sigma\lambda}= - \bar v_\lambda u_\sigma$.
Thus,  the transformation matrix from the common-used basis to the helicity basis is
\begin{equation}
\pmatrix{u_\sigma\cr
v_\sigma\cr}={\cal U} \pmatrix{u_\lambda\cr
v_\lambda\cr},\,\,\quad{\cal U} = \pmatrix{A&B\cr
B&A}
\end{equation}
Neither $A$ nor $B$ are unitary:
\begin{eqnarray}
A= (a_{++} +a_{+-}) (\sigma_\mu a^\mu) +(-a_{-+} +a_{--})
(\sigma_\mu a^\mu) \sigma_3\,,\\
B= (-a_{++} +a_{+-}) (\sigma_\mu a^\mu) +(a_{-+} +a_{--})
(\sigma_\mu a^\mu) \sigma_3\,,
\end{eqnarray}
where
\begin{eqnarray}
a^0 &=& -i\cos (\theta/2) \sin (\phi/2) \in \Im m\,,\quad
a^1 = \sin (\theta/2) \cos (\phi/2)\in \Re e\,,\\
a^2 &=& \sin (\theta/2) \sin (\phi/2) \in \Re e\,,\quad
a^3 = \cos (\theta/2) \cos (\phi/2)\in \Re e\,,
\end{eqnarray}
and
\begin{eqnarray}
a_{++} &=&\frac{\sqrt{(E+m)(E+p)}}{2\sqrt{2} m}\,,\quad
a_{+-} =\frac{\sqrt{(E+m)(E-p)}}{2\sqrt{2} m}\,,\\
a_{-+} &=&\frac{\sqrt{(E-m)(E+p)}}{2\sqrt{2} m}\,,\quad
a_{--} =\frac{\sqrt{(E-m)(E-p)}}{2\sqrt{2} m}\,.
\end{eqnarray}
However, $A^\dagger A+B^\dagger B =\openone$, so the matrix ${\cal U}$
is unitary. Please note that the $4\times 4$  matrix acts
on the {\it spin}  indices ($\sigma$,$\lambda$), and does not on
the spinorial indices. Alternatively,
the transformation can be written:
\begin{eqnarray}
u_\sigma^\alpha &=& [A_{\sigma\lambda}\otimes I_{\alpha\beta}
+B_{\sigma\lambda}\otimes \gamma^5_{\alpha\beta}] u_\lambda^\beta\,,\\
v_\sigma^\alpha &=& [A_{\sigma\lambda}\otimes I_{\alpha\beta}
+B_{\sigma\lambda}\otimes \gamma^5_{\alpha\beta}] v_\lambda^\beta\,.
\end{eqnarray}

We now investigate the properties of the helicity-basis 4-spinors
with respect to the discrete symmetry operations $P$ and $C$.
It is expected that $\lambda\rightarrow -\lambda$ under parity,
as Berestetski\u{\i}, Lifshitz and Pitaevski\u{\i}
claimed~\cite{Lan}.\footnote{Indeed, if ${\bf x}\rightarrow -{\bf x}$,
then the vector ${\bf p}\rightarrow -{\bf p}$, but the axial vector
${\bf S}\rightarrow {\bf S}$, that implies the above statement.}
With respect to ${\bf p} \rightarrow -{\bf p}$ (i.~e.,
the spherical system  angles $\theta \rightarrow \pi-\theta$,
$\phi \rightarrow \pi+\phi$) the helicity 2-eigenspinors
transform as follows: $\phi_{\uparrow\downarrow} \Rightarrow
-i \phi_{\downarrow\uparrow}$, ref.~\cite{Dv1}.
Hence,
\begin{eqnarray}
Pu_\uparrow (-{\bf p}) &=& -i u_\downarrow ({\bf p})\,,
Pv_\uparrow (-{\bf p}) = +i v_\downarrow ({\bf p})\,,\\
Pu_\downarrow (-{\bf p}) &=& -i u_\uparrow ({\bf p})\,,
Pv_\downarrow (-{\bf p}) = +i v_\uparrow ({\bf p})\,.
\end{eqnarray}
Thus, on the level of classical fields, we observe that
the helicity 4-spinors transform to the 4-spinors of the opposite
helicity.

The charge conjugation operation is defined as
\begin{equation}
C =\pmatrix{0&\Theta\cr
-\Theta & 0\cr} {\cal K}\,.\label{chcon}
\end{equation}
Hence, we observe
\begin{eqnarray}
Cu_\uparrow ({\bf p}) &=& - v_\downarrow ({\bf p})\,,
Cv_\uparrow ({\bf p}) = +  u_\downarrow ({\bf p})\,,\\
Cu_\downarrow ({\bf p}) &=& + v_\uparrow ({\bf p})\,,
Cv_\downarrow ({\bf p}) = - u_\uparrow ({\bf p})\,.
\end{eqnarray}
due to the properties of the Wigner operator $\Theta \phi_\uparrow^\ast =
-\phi_\downarrow$ and $\Theta \phi_\downarrow^\ast = +\phi_\uparrow$.
For the $CP$ (and $PC$) operation we get:
\begin{eqnarray}
C P u_\uparrow (-{\bf p}) &=& -PC u_\uparrow (-{\bf p})
= +i v_\uparrow ({\bf p})\,,\\
C P u_\downarrow
(- {\bf p}) &=& - P C u_\downarrow (-{\bf p}) = -i v_\downarrow ({\bf
p})\,,\\
C P v_\uparrow (-{\bf p}) &=& - P C v_\uparrow (-{\bf p}) =
+  i u_\uparrow ({\bf p})\,,\\
C P v_\downarrow (-{\bf p}) &=& - P C v_\downarrow (-{\bf p}) =
- i u_\downarrow ({\bf p})\,,
\end{eqnarray}
which are different from the Dirac `common-used' case.

Similar conclusions can be drawn in the Fock space.

\section{The $(1,0)\oplus (0,1)$ case.}

In this Section we are going to investigate the behavious of the field functions of the $(1,0)\oplus (0,1)$ representation in the helicity basis with respect to $P$, $C$ and $CP$ operations. 

Let us start from the Klein-Gordon equation written for the 3-component function ($\hbar=c=1$):
\begin{equation}
(E^2-{\bf p}^2)\psi_{(3)}=m^2\psi_{(3)}. \label{1b}
\end{equation}
The function $\psi$ describe the particles, which is usually referred as spin 1; we refer to it as a ``3-spinor". On choosing the basis where ${\bf S}^i_{jk} = -i\epsilon^{ijk}$ one can derive the following property for any 3-vector ${\bf a}$: 
\begin{equation}
({\bf S\cdot a})^2_{ij}={\bf a}^2\delta_{ij}-a^i a^j
\end{equation} 
Then the equation (\ref{1b}) can be re-written in the form:
\begin{equation}
( E-{\bf S\cdot p})(E+{\bf S\cdot p})_{ij}{\bbox\psi}^j- p_i p_j{\bbox\psi}^j=m^2\psi^i.
\label{2b}
\end{equation}
In the coordinate space it is of the second order in the time derivative, 
but as in the spin-1/2 case~\cite{valerihelic} we can reduce it to the set of  the 3-``spinor" equations of the first orders. The procedure permits us to consider the hamiltonian-like form $i\hbar {\partial \psi \over \partial t}=\widehat{H}\psi$ and make it easier to find the energy eigenstates of the problem.

We can denote:
\begin{eqnarray}
(E+{\bf S\cdot p}){\bbox \psi}=m{\bbox\xi} \label{a}\\
p^ip^j\psi^j={\bf p\,(p}\cdot{\bbox\psi})=m {\bf p}\,\varphi . \label{b}
\end{eqnarray}
Hence the equation (\ref{2b}) is written
\begin{equation}
m(E-{\bf S\cdot p}){\bbox\xi}-m{\bf p}\,\varphi=m^2{\bbox\psi}. \label{c}
\end{equation}
Now, we insert the properties 
\begin{equation}
({\bf S\cdot p})^{ij}{\bbox\psi}^j=({\bbox\nabla}\times{\bbox\psi})^i, \hspace{5mm} p^i p^j {\bbox\psi}^j=-[{\bbox\nabla}\left({\bbox\nabla}\cdot{\bbox\psi}\right)]^i, \label{3}
\end{equation}
and define ${\bbox\psi}={\bf E}-i{\bf B}$. We can obtain (cf. with ref.~\cite{Dv2a}) 
\begin{equation}
 \nabla\times{\bf B}-\frac{\partial {\bf E}}{\partial t}=-m\cdot{\sf Im}({\bbox\xi}), \hspace{5mm}
\nabla\times{\bf E}+\frac{\partial {\bf B}}{\partial t}=m\cdot{\sf Re}({\bbox \xi}),
\end{equation}
and
\begin{equation}
\nabla\cdot{\bf B}=-m\cdot{\sf Re}(\varphi)+const_x\,, \hspace{5mm}
\nabla\cdot{\bf E}=-m\cdot{\sf Im}(\varphi)+const_x\,,
\end{equation}
respectively by means of separation of the equations (\ref{a},\ref{b}) into the real and imaginary parts. Next, we fix $\varphi=im\phi$ and ${\bbox\xi}=im{\bf A}$, with $\phi$ and ${\bf A}$ being the electromagnetic-like potentials. The well-known Proca equation follows:
\begin{equation}
\partial_{\mu}F^{\mu \nu}+m^2 A^{\nu}=0.
\label{proca}
\end{equation}
For the sake of completeness let su substitute  $\varphi$ and ${\bbox \xi}$ in the equation (\ref{c}). The result is $-\frac{\partial {\bf A}}{\partial t}-\nabla \phi={\bf E}$ y $\nabla \times {\bf A}={\bf B}$, that is equivalent to the second Proca equation:
\begin{equation}
F^{\mu \nu}=\partial^{\mu}A^{\nu}-\partial^{\nu}A^{\mu}\,.\label{sPro}
\end{equation}

Let us take the complex conjugates of the equations (\ref{a}), (\ref{b},\ref{c}) and now define  ${\bbox \chi}={\bf E}+i{\bf B}$. As a result we have 
\begin{eqnarray}
(E-{\bf S\cdot p}){\bbox\chi}=-m{\bbox\xi}&\hspace{3mm}\mbox{or}\hspace{3mm}&
(E-{\bf S\cdot p})({\bf E}+i{\bf B})=-im^2{\bf A}, \label{aa}\\
p^ip^j\chi^j={\bf p(p}\cdot{\bbox\chi})=-m{\bf p}\varphi&\hspace{3mm}\mbox{or}\hspace{3mm}& {\bf p\, [p}\cdot({\bf E}+i{\bf B})]=-im^2{\bf p}\phi, \label{bb}\\
(E+{\bf S\cdot p}){\bbox\xi}-{\bf p}\varphi=-m{\bbox\chi}&\hspace{3mm}\mbox{or}\hspace{3mm}&
(E+{\bf S\cdot p}){\bf A}-{\bf p}\phi=i({\bf E}+i{\bf B}), \label{cc}
\end{eqnarray}
It is possible to re-write the above equations (and their complex conjugates)
in the $10\times 10$ matrix equation (with appropriate choice of matrices) for spin-1 particles (cf.~\cite{greiner}).
\begin{equation}
\pmatrix{
0&0&0& 0&0&0 &-E&ip_z&-ip_y &p_x\cr
0&0&0& 0&0&0 &-ip_z&-E&ip_x &p_y\cr
0&0&0& 0&0&0 &ip_y&-ip_x&-E &p_z\cr
0&0&0& 0&0&0 &E&ip_z&-ip_y &-p_x\cr
0&0&0& 0&0&0 &-ip_z&E&ip_x &-p_y\cr
0&0&0& 0&0&0 &ip_y&-ip_x&E &-p_z\cr
-E&-ip_z&ip_y &0&0&0& 0&0&0 &0\cr
ip_z&-E&-ip_x &0&0&0& 0&0&0 &0\cr
-ip_y&ip_x&-E &0&0&0& 0&0&0 &0\cr
-p_x&-p_y&-p_z &0&0&0& 0&0&0 &0}
\pmatrix{
\chi_1\cr \chi_2\cr \chi_3\cr
\psi_1\cr \psi_2\cr \psi_3\cr
\xi_1\cr \xi_2\cr \xi_3\cr
\varphi}=
m\pmatrix{\chi_1\cr \chi_2\cr \chi_3\cr
\psi_1\cr \psi_2\cr \psi_3\cr
\xi_1\cr \xi_2\cr \xi_3\cr
\varphi},\label{kemmer}\end{equation}
which is in the symbolic form:
\begin{equation}
\left( \begin{array}{cccc}
0_{3\times 3} & 0_{3\times 3} & -(E+{\bf S\cdot p})_{3\times 3} & {\bf p}_{3\times 1}\\
0_{3\times 3} & 0_{3\times 3} & (E-{\bf S\cdot p})_{3\times 3} & -{\bf p}_{3\times 1}\\
-(E-{\bf S\cdot p})_{3\times 3} & 0_{3\times 3} & 0_{3\times 3}& 0_{3\times 1}\\
-{\bf p}_{1\times 3} & 0_{1\times 3} & 0_{1\times 3} & 0 \end{array} \right) \Xi=m\Xi\,\label{dkp}
\end{equation}
for the 10-component field function $\Xi=\mbox{column} (\vec{\chi}, \vec{\psi}, \vec{\xi}, \varphi)$. This first-order equation is known as the Duffin-Kemmer-Petiau (DKP) equation~\cite{greiner}. It contains the part corresponding to the 4-vector potential~\footnote{It would be of interest to research the helicity basis for the DKP equation, as we did for the Dirac  equation. However, we leave this task for the future works. Instead, we are going to consider the helicity basis of the solutions of the Weinberg-Tucker-Hammer second-order equations below.} At first sight, for the construction of  (\ref{kemmer}) we have used the equations (\ref{c}) and (\ref{aa}-\ref{cc}), and omitted the equations (\ref{a}, \ref{b}). However, one can show that our DKP equation contains that information. If we write (\ref{a}-\ref{c}) and (\ref{aa}) in the matrix form, we can also write
\begin{equation}
\pmatrix{(E+{\bf S\cdot p})_{3\times 3}&0_{3\times 3}&0_{3\times 1}&-m_{3\times 3}\cr 0_{3\times 3}&(E-{\bf S\cdot p})_{3\times 3}&0_{3\times 1}&m_{3\times 3}\cr 
{\bf p}_{1\times 3}&0_{1\times 3}&-m_{1\times 1}&0_{1\times 3}\cr
-m_{3\times 3}&0_{3\times 3}&-{\bf p}_{3\times 1}&(E-{\bf S\cdot p})_{3\times 3}}\pmatrix{\vec{\psi}\cr \vec{\chi}\cr \varphi\cr \vec{\xi}}=0\,,\label{kemmer2} \end{equation}
which is related to (\ref{dkp}).
It is more convenient to write this equation in terms of ${\bf E}$, ${\bf B}$, $\phi$ and ${\bf A}$. We use the unitary transformation with 
\begin{equation}
{\cal V}=\frac{1}{\sqrt{2}} 
\pmatrix{1_{3\times 3}&1_{3\times 3}&0_{3\times 1}&0_{3\times 3}\cr i_{3\times 3}&-i_{3\times 3}&0_{3\times 1}&0_{3\times 3}\cr 0_{1\times 3}&0_{1\times 3}&2_{1\times 1}&0_{1\times 3}\cr 0_{3\times 3}&0_{3\times 3}&0_{3\times 1}&2_{3\times 3}}.
\end{equation}
As a result we have
\begin{equation}
\pmatrix{E_{3\times 3}&-i({\bf S\cdot p})_{3\times 3}&0_{3\times 1}&0_{3\times 3}\cr i({\bf S\cdot p})_{3\times 3}&E_{3\times 3}&0_{3\times 1}&-2im_{3\times 3}\cr {\bf p}_{1\times 3}&-i{\bf p}_{1\times 3}&-2m_{1\times 1}&0_{1\times 3}\cr 
-m_{3\times 3}&im_{3\times 3}&-2{\bf p}_{3\times 1}&2(E-{\bf S\cdot p})_{3\times 3}} \pmatrix{{\bf E}\cr {\bf B}\cr im\phi\cr im{\bf A}},
\label{kemmer3}\end{equation}
where ${\bf p}_{1\times 3}=(p_x,p_y,p_z)$ is the row and ${\bf p}_{3\times 1}$ is the column. It is equivalent to the Proca set.

Taking into account the Proca equations (\ref{proca},\ref{sPro}), the deinitions of ${\bf E}^i= F^{i0}$, ${\bf B}^i=-{1\over 2}\epsilon^{ijk} F^{jk}$ and the definition of the Levi-Civita tensor, we can obtain the Tucker-Hammer equation~\cite{tucker}:
\begin{equation} \left(\begin{array}{cc}
E^2-{\bf p}^2-2m^2&E^2-{\bf p}^2+2E({\bf S\cdot p})+2({\bf S\cdot p})^2\\
E^2-{\bf p}^2-2E({\bf S\cdot p})+2({\bf S\cdot p})^2&E^2-{\bf p}^2-2m^2
\end{array}\right) \left( \begin{array}{c}
\chi\\ \psi
\end{array}\right)=0. 
\label{tucker1}
\end{equation}
In the covariant form the equation (\ref{tucker1}) is written:
\begin{equation}
\left( \gamma^{\mu \nu}p_{\mu}p_{\nu}+p^{\mu}p_{\mu}-2m^2\right)\Psi_{(6)}(p^{\mu})=0. 
\end{equation}
with the $6\times 6$ matrices~\cite{barut})
\[
\gamma^{00}=\left(\begin{array}{cc}0&1_{3\times 3}\\1_{3\times 3}&0\end{array}\right),\hspace{5mm}
\gamma^{i0}=\gamma^{0i}=\left(\begin{array}{cc}0&-S^i\\S^i&0\end{array}\right),\]
\begin{equation}
\gamma^{ij}=\left(\begin{array}{cc}0&-\delta_{ij}+S_iS_j+S_jS_i\\-\delta_{ij}+S_iS_j+S_jS_i&0\end{array}\right)\,.
\label{gamma}
\end{equation}
In the coordinate space we have:
\begin{equation}
\left( \gamma^{\mu \nu}\partial_{\mu}\partial_{\nu}+\partial^{\mu}\partial_{\mu}+2m^2\right)\Psi(x^{\mu})=0\,. \label{tucker2}
\end{equation}

If we set the condition $\partial_{\mu}\partial_{\mu}\rightarrow -m^2$ we can recover the Weinberg equation, ref.~\cite{weinbergp}\footnote{We should mention that this procedure is not quite clear, because the dispersion relations of the Weinberg equation and the Tucker-Hammer equation may be different (see~\cite{valeri7}). The Weinberg equation permits, in general, the tachyonic solutions, $E^2-{\bf p}^2=-m^2$.}:
\begin{equation} \Gamma\pmatrix{\chi\cr\psi}=\left(\begin{array}{cc}
-m^2&m^2+2E({\bf S\cdot p})+2({\bf S\cdot p})^2\\
m^2-2E({\bf S\cdot p})+2({\bf S\cdot p})^2&-m^2
\end{array}\right) \left( \begin{array}{c}
 \chi\\
 \psi
\end{array}\right)=0\,, 
\label{weinberg}
\end{equation}
which is in the covariant form
\begin{equation}
(\gamma^{\mu \nu}\partial_{\mu}\partial_{\nu}+ m^2)\Psi(x^{\mu})=0\,.
\label{ecwein} 
\end{equation}
Thus, from what we have seen above, we can conclude that the Duffin-Kemmer-Petiau, Proca, Weinberg and Tucker-Hammer equations are all related one another. They can be obtained by  various transformations from the relativistic dispersion relation, $E^2-{\bf p}^2=m^2$.

Let us consider the equation (\ref{tucker1}) as a set of equations for the bivector components in the helicity basis. Then, we have ($p=\mid {\bf p} \mid $):
\begin{eqnarray}
(E^2-p^2+2Ep+2p^2)\psi_{\uparrow}&=&(2m^2-(E^2-p^2))\chi_{\uparrow},\nonumber\\
(E^2-p^2-2Ep+2p^2)\chi_{\uparrow}&=&(2m^2-(E^2-p^2))\psi_{\uparrow},
\hspace{3mm}(h=1)\label{relacionhelic1}\\
(E^2-p^2-2Ep+2p^2)\psi_{\downarrow}&=&(2m^2-(E^2-p^2))\chi_{\downarrow},\nonumber\\ 
(E^2-p^2+2Ep+2p^2)\chi_{\downarrow}&=&(2m^2-(E^2-p^2))\psi_{\downarrow},
\hspace{3mm}(h=-1)\label{relacionhelic2}\\
(E^2-p^2)\psi_{\rightarrow}&=&(2m^2-(E^2-p^2))\chi_{\rightarrow},\nonumber\\ 
(E^2-p^2)\chi_{\rightarrow}&=&(2m^2-(E^2-p^2))\psi_{\rightarrow},
\hspace{3mm}(h=0)\label{relacionhelic3}.
\end{eqnarray}
where the 3-``spinors" are in the helicity basis  (see~\cite[p.192]{Var}):
\begin{equation}
\chi_{\uparrow}=\left(\begin{array}{c}
\frac{1+\cos\theta}{2}e^{-i\phi}\\
\frac{\sin\theta}{\sqrt{2}}\\
\frac{1-\cos\theta}{2}e^{i\phi}\end{array}\right),\hspace{2mm}
\chi_{\rightarrow}=\left(\begin{array}{c}
-\frac{\sin\theta}{\sqrt{2}}e^{-i\phi}\\
\cos\theta\\
\frac{\sin\theta}{\sqrt{2}}e^{i\phi}\end{array}\right),\hspace{2mm}
\chi_{\downarrow}=\left(\begin{array}{c}
\frac{1-\cos\theta}{2}e^{-i\phi}\\
-\frac{\sin\theta}{\sqrt{2}}\\
\frac{1+\cos\theta}{2}e^{i\phi}\,.\end{array}\right)
\end{equation}
The normalization condition is chosen $\chi^{\dagger}\chi=1$.

Taking into account (\ref{relacionhelic1}-\ref{relacionhelic3}) we can write the bivectors $u_{\uparrow, \downarrow, \rightarrow}=\pmatrix{\chi_{\uparrow, \downarrow, \rightarrow}\cr \psi_{\uparrow, \downarrow, \rightarrow}}$ in the following way:
\begin{equation}
u_{1,\uparrow}=N_{\uparrow}\pmatrix{\chi_{\uparrow}\cr \frac{2m^2-(E^2-p^2)}{E^2-p^2+2Ep+2p^2}\chi_{\uparrow}},\hspace{2mm}
u_{1,\rightarrow}=N_{\rightarrow}\pmatrix{\chi_{\rightarrow}\cr \frac{2m^2-(E^2-p^2)}{E^2-p^2}\chi_{\rightarrow}},\hspace{2mm}
u_{1,\downarrow}=N_{\downarrow}\pmatrix{\chi_{\downarrow}\cr \frac{2m^2-(E^2-p^2)}{E^2-p^2-2Ep+2p^2}\chi_{\downarrow}}\,.
\end{equation}

Let us now introduce $\overline{u}_{\lambda}=u^{\dagger}\gamma^{00}$, $v_{\lambda}=\gamma^5 u_{\lambda}$ (where $\gamma^5=\pmatrix{1_{3\times 3}&0_{3\times 3}\cr 0_{3\times 3}&-1_{3\times 3}}$). After the  normalization to the unit and imposing $m^2=E^2-p^2$, our bivectors are then
\begin{equation}
u_{1,\uparrow}=\frac{1}{\sqrt{2}}\left(\begin{array}{c} 
\frac{E+p}{m}\chi_{\uparrow}\\
\frac{m}{E+p}\chi_{\uparrow}\end{array}\right),\,
u_{1,\rightarrow}=\frac{1}{\sqrt{2}}\left(\begin{array}{c} 
\chi_{\rightarrow}\\
\chi_{\rightarrow}\end{array}\right),\,
u_{1,\downarrow}=\frac{1}{\sqrt{2}}\left(\begin{array}{c} 
\frac{m}{E+p}\chi_{\downarrow}\\
\frac{E+p}{m}\chi{\downarrow}\end{array}\right),
\end{equation}
\begin{equation}
v_{1,\uparrow}=\frac{1}{\sqrt{2}}\left(\begin{array}{c} 
\frac{E+p}{m}\chi_{\uparrow}\\
-\frac{m}{E+p}\chi_{\uparrow}\end{array}\right),\,
v_{1,\rightarrow}=\frac{1}{\sqrt{2}}\left(\begin{array}{c} 
\chi_{\rightarrow}\\
-\chi_{\rightarrow}\end{array}\right),\,
v_{1,\downarrow}=\frac{1}{\sqrt{2}}\left(\begin{array}{c} 
\frac{m}{E+p}\chi_{\downarrow}\\
-\frac{E+p}{m}\chi{\downarrow}\end{array}\right).
\end{equation}

Now we study the discrete symmetry operations for spin-1 case (as we did for spin-1/2 case in the previous Section). The bivectors have the following properties:
\begin{enumerate} 
\item 
The Parity (${\bf p}\rightarrow -{\bf p}$, $\theta\rightarrow \pi-\theta$, $\phi\rightarrow \pi +\phi$). We note that the 3-``spinors"  are transformed as $\chi_h\rightarrow-\chi_{-h}$; the parity operator is $P=\gamma^{00}$ (it is analogous to that which was used for spin-1/2 (see (\ref{par})). Therefore,
\begin{equation}
Pu_{1,\uparrow}(-{\bf p})=-u_{1,\downarrow}({\bf p}), \hspace{2mm} 
Pu_{1,\rightarrow}(-{\bf p})=-u_{1,\rightarrow}({\bf p}), \hspace{2mm}
Pu_{1,\downarrow}(-{\bf p})=-u_{1,\uparrow}({\bf p})\,.
\end{equation}
And,
\begin{equation}
Pv_{1,\uparrow}(-{\bf p})=+v_{1,\downarrow}({\bf p}), \hspace{2mm} 
Pv_{1,\rightarrow}(-{\bf p})=+v_{1,\rightarrow}({\bf p}), \hspace{2mm}
Pv_{1,\downarrow}(-{\bf p})=+v_{1,\uparrow}({\bf p})\,.
\end{equation}

\item 
The Charge Conjugation is defined
\begin{equation}C=e^{i\alpha}\pmatrix{0&\Theta\cr-\Theta&0}{\cal K}\end{equation}
(similarly to (\ref{chcon})) with $\Theta_{[1]}=\left(\begin{array}{ccc}0&0&1\\0&-1&0\\1&0&0\end{array}\right)$.
Hence, $\Theta\chi_{\uparrow}^\ast=\chi_{\downarrow}$, $\Theta\chi_{\downarrow}^\ast=\chi_{\uparrow}$, $\Theta\chi_{\rightarrow}^\ast=-\chi_{\rightarrow}$. Finally, we have
\begin{equation}
Cu_{1,\uparrow}({\bf p})=+e^{i\alpha}v_{1,\downarrow}({\bf p}), \hspace{2mm} 
Cu_{1,\rightarrow}({\bf p})=-e^{i\alpha}v_{1,\rightarrow}({\bf p}), \hspace{2mm}
Cu_{1,\downarrow}({\bf p})=+e^{i\alpha}v_{1,\uparrow}({\bf p})\,.
\end{equation}
And
\begin{equation}
Cv_{1,\uparrow}({\bf p})=-e^{i\alpha}u_{1,\downarrow}({\bf p}), \hspace{2mm} 
Cv_{1,\rightarrow}({\bf p})=+e^{i\alpha}u_{1,\rightarrow}({\bf p}), \hspace{2mm}
Cv_{1,\downarrow}({\bf p})=-e^{i\alpha}u_{1,\uparrow}({\bf p})\,.
\end{equation}

\item The $CP$ and $PC$ Operations:
\begin{equation} CPu_{1,\uparrow}({\bf -p})=-PCu_{1,\uparrow}({\bf -p})=-e^{i\alpha}v_{1,\uparrow}({\bf p}),\label{cp1} \end{equation}
\begin{equation}CPv_{1,\uparrow}({\bf -p})=-PCv_{1,\uparrow}({\bf -p})=-e^{i\alpha}u_{1,\uparrow}({\bf p}),\end{equation}
\begin{equation}CPu_{1,\downarrow}({\bf -p})=-PCu_{1,\downarrow}({\bf -p})=-e^{i\alpha}v_{1,\downarrow}({\bf p}), \end{equation} 
\begin{equation}CPv_{1,\downarrow}({\bf -p})=-PCv_{1,\downarrow}({\bf -p})=-e^{i\alpha}u_{1,\downarrow}({\bf p}),\end{equation}
\begin{equation}CPu_{1,\rightarrow}({\bf -p})=-PCu_{1,\rightarrow}({\bf -p})=+e^{i\alpha}v_{1,\rightarrow}({\bf p}),\end{equation}
\begin{equation}CPv_{1,\rightarrow}({\bf -p})=-PCv_{1,\rightarrow}({\bf -p})=+e^{i\alpha}u_{1,\rightarrow}({\bf p}).\label{cp2}\end{equation}

\end{enumerate}

We found within the classical field theory that the properties of particle and anti-particle of spin-1 are different comparing with the known cases (when the basis is chosen in such a way that the solutions are the eigenstates of the parity).

\section{The Conclusions.}

\begin{itemize}

\item
Similarly to the $({1\over 2},{1\over 2})$ representation~\cite{Grei},
the $({1\over 2},0)\oplus (0,{1\over 2})$ and $(1,0)\oplus (0,1)$ field functions
in the helicity basis are {\it not} eigenstates of the
common-used parity operator; $\vert {\bf p},\lambda> \Rightarrow
\vert -{\bf p},-\lambda >$  on the classical level. This
is in accordance with the earlier consideration of Berestetski\u{\i},
Lifshitz and Pitaevski\u{\i}.

\item
Helicity field functions may satisfy the ordinary Dirac equation
with $\gamma$'s to be in the spinorial representation. 

\item
Helicity field functions can be expanded in the set of the Dirac
4-spinors by means of the matrix ${\cal U}^{-1}$ given in this paper.

\item
$P$ and $C$ operations anticommute  in this framework on the
classical level.

\item
The different formulations of the spin-1 particles are all connected by algebraic transformations.

\item
The properties of spin-1 solutions in the helicity basis with respect to $P$, $C$, $CP$ are similar to those in the spin-1/2 case.

\end{itemize}

In order to make the above conclusions to be more firm one should repeat the calculations in the Fock space within the ``secondary quantization" framework 
(see~\cite{valerihelic} for the spin-1/2 case).


\begin{thebibliography}{99}

\bibitem{Barut} G. Ziino, Ann. Fond. Broglie {\bf 14} (1989) 427; ibid
{\bf 16} (1991) 343; A.  Barut and G.  Ziino, Mod.  Phys.  Lett.  A{\bf
8} (1993) 1011; G. Ziino, Int. J. Mod. Phys. A{\bf 11} (1996) 2081.

\bibitem{DasG} N. D. S. Gupta, Nucl. Phys. B{\bf 4} (1967) 147;
D. V. Ahluwalia, Int. J. Mod. Phys. A{\bf 11}
(1996) 1855; V. Dvoeglazov, Hadronic J. {\bf 20} (1997) 435.

\bibitem{Dv2} V. V. Dvoeglazov, Mod. Phys. Lett. A{\bf 12} (1997) 2741.

\bibitem{Dv2a} V. V. Dvoeglazov, Spacetime and Substance {\bf 3}(12)
(2002) 28; Rev. Mex. Fis., Supl. 1,  {\bf 49} (2003) 99.

\bibitem{Dv3} V. V. Dvoeglazov, Physica Scripta {\bf 64} (2001) 201.

\bibitem{Dv4} V. V. Dvoeglazov, Hadronic J. {\bf 25} (2002) 137.

\bibitem{Dv5} V. V. Dvoeglazov, Hadronic J. {\bf 26} (2003) 299.

\bibitem{SG} A. Sankaranarayanan and R. H. Good, jr., Nuovo Cim. {\bf 36}
(1965) 1303.

\bibitem{Dva} D. V. Ahluwalia, M. B. Johnson and T. Goldman, Phys. Lett.
B{\bf 316} (1993) 102; V. V. Dvoeglazov, Int. J. Theor. Phys. {\bf 37},
(1998) 1915, and references therein.

\bibitem{Lan} V. B. Berestetski\u{\i}, E. M. Lifshitz and
L. P. Pitaevski\u{\i}, {\it Quantum Electrodynamics} (Pergamon Press, 1982),
\S 16.

\bibitem{Novozh} Yu. V. Novozhilov, {\it Introduction to Elementary Particle
Physics} (Pergamon Press, 1975), \S 4.3, 6.2.

\bibitem{Grei} H. M. Ruck y W. Greiner, J. Phys. G: Nucl. Phys. {\bf 3} (1977) 657.

\bibitem{JW} M. Jackob and G. C. Wick, Ann. Phys. {\bf 7} (1959) 404.

\bibitem{Var} D. A. Varshalovich, A. N. Moskalev and V. K. Khersonski\u{\i},
{\it Quantum Theory of Angular Momentum} (World Scientific, 1988), \S 6.2.5.

\bibitem{Dv1} V. V. Dvoeglazov, Fizika B{\bf 6} (1997) 111.

\bibitem{Gerst} J. J. Sakurai, {\it Advanced Quantum Mechanics.} (Addison-Wesley, 1967), \S 3.2; A. Gersten, Found. Phys. Lett. {\bf 12} (1999) 291; ibid. {\bf 13} (2000) 185; V. V. Dvoeglazov, J. Phys. A: Math. Gen.  {\bf 33} (2000) 5011.

\bibitem{valerihelic} V. V. Dvoeglazov, In  {\it Memorias de la 8a Reuni\'on Nacional Acad\'emica de F\'{\i}sica y Matem\'aticas, 12-16 de Mayo de 2003, ESFM-IPN, M\'exico, D.F.}, p. 45-54.

\bibitem{tucker} R. H. Tucker y C. L. Hammer,  Phys. Rev. D{\bf 3} (1971) 2448.

\bibitem{barut} A. O. Barut, I. Muzinich and D. Williams, Phys. Rev. {\bf 130} (1963) 442.

\bibitem{weinbergp} S. Weinberg, Phys. Rev. {\bf 133}B (1964) 1318.

\bibitem{valeri7} V. V. Dvoeglazov, Helv. Phys. Acta, {\bf 70} (1997) 677.

\bibitem{greiner} W. Greiner, {\it Relativistic Quantum Mechanics.} The 1st English Ed. (Springer, 1990).


\end{thebibliography}
\end{document}